\documentclass[aps,pra,twocolumn,superscriptaddress,showpacs]{revtex4-1}

\usepackage{graphicx}% Include figure files

\usepackage{color}

\usepackage[mathlines]{lineno}

\usepackage{amsfonts}
\usepackage{amsmath}

\def\Tr{\operatorname{Tr}}

\let\originalleft\left
\let\originalright\right

\renewcommand{\left}{\mathopen{}\mathclose\bgroup\originalleft}
\renewcommand{\right}{\aftergroup\egroup\originalright}
\newcommand{\abs} [1]{\ensuremath{\left|#1\right|}}

\newcommand{\bra}[1]{\ensuremath{\left\langle#1\right|}}
\newcommand{\ket}[1]{\ensuremath{\left|#1\right\rangle}}
\newcommand{\bracket}[2]{\ensuremath{\left\langle#1 \vphantom{#2}\right| \left. #2 \vphantom{#1}\right\rangle}}

\begin{document}

%\preprint{}

\title{Preserving photon qubits in an unknown quantum state with Knill Dynamical Decoupling --- Towards an all optical quantum memory}

\author{Manish K. Gupta}
\email{\color{magenta}mgupta3@lsu.edu}
\affiliation{Hearne Institute for Theoretical Physics, Department of Physics and Astronomy, Louisiana State University, Baton Rouge, Louisiana 70803, USA.}
\author{Erik J. Navarro}
\affiliation{Department of Physics, California State University, Chico, Chico, CA 95929-0202, USA.}
\author{Todd A. Moulder}
\affiliation{Hearne Institute for Theoretical Physics, Department of Physics and Astronomy, Louisiana State University, Baton Rouge, Louisiana 70803, USA.}
\author{Jason D. Mueller}
\affiliation{Hearne Institute for Theoretical Physics, Department of Physics and Astronomy, Louisiana State University, Baton Rouge, Louisiana 70803, USA.}
\author{Ashkan Balouchi}
\affiliation{Hearne Institute for Theoretical Physics, Department of Physics and Astronomy, Louisiana State University, Baton Rouge, Louisiana 70803, USA.}
\author{Katherine L. Brown}
\affiliation{Hearne Institute for Theoretical Physics, Department of Physics and Astronomy, Louisiana State University, Baton Rouge, Louisiana 70803, USA.}
\author{Hwang Lee}
\affiliation{Hearne Institute for Theoretical Physics, Department of Physics and Astronomy, Louisiana State University, Baton Rouge, Louisiana 70803, USA.}
\author{Jonathan P. Dowling}
\affiliation{Hearne Institute for Theoretical Physics, Department of Physics and Astronomy, Louisiana State University, Baton Rouge, Louisiana 70803, USA.}

%\thanks{}


\date{\today}

\begin{abstract}
The implementation of polarization-based quantum communication is limited by signal loss and decoherence caused by the birefringence of a single-mode fiber. We investigate the Knill dynamical decoupling scheme, implemented using half-wave plates, to minimize decoherence and show that a fidelity greater than $99\%$ can be achieved in absence of rotation error and fidelity greater than $96\%$ can be achieved in presence of rotation error.  Such a scheme can be used to preserve any quantum state with high fidelity and has potential application for constructing all optical quantum delay line, quantum memory, and quantum repeater.
\end{abstract}

%\keywords{}

\maketitle

\section{Introduction}
\label{sec:Introduction}

The storage of quantum states and its distribution over long distances is essential for emerging quantum technologies such as quantum networks and long distance quantum cryptography. Quantum networks and quantum cryptographic systems \cite{BB84,RevModPhys.74.145} use photonic qubits as information carriers because of their quantum nature and low-loss coefficient pertaining to transmission, both in optical fiber \cite{Stucki2009} and in free space \cite{Ursin2007}. Nevertheless, the losses become significant when we envision transmission over hundreds of kilometers or more and is accompanied by decoherence. In classical telecommunication the issue of signal loss is overcome by use of amplifiers (also called repeaters) that amplify the signals. Unfortunately this option of signal amplification is forbidden in quantum communication because of the no-cloning theorem \cite{Dieks1982271,wootters1982} unless one restricts oneself to an orthogonal set of states; in fact a quantum protocol such as quantum key distribution arises precisely due to existence of non-orthogonal states.

The problem of signal amplification in quantum communication can be overcome with the use of sophisticated entanglement-based scheme known as a quantum repeater \cite{PhysRevLett.81.5932}. Entanglement has a special property that it can be swapped. Given a entangled state between $A$ and $B$ and another between $C$ and $D$, it is possible to create entanglement between $A$ and $D$ by performing a joint measurement in the basis of entangled states, followed by classical communication of the result to the location $A$ and/or $D$. The states at $A$ and $D$ need to be preserved in a quantum memory for the period of time in which the classical result are communicated for the teleportation protocol to be successful \cite{RevModPhys.83.33}. Current schemes \cite{PhysRevLett.96.240501,PhysRevA.85.032313,PhysRevLett.81.5932} suggest use of atomic states to preserve the state of qubit, but it would be beneficial if the qubit state can be preserved in an all optical quantum memory so that all optical implementation of quantum repeater device is feasible \cite{nature07241}.

Another issue faced by any quantum repeater scheme\cite{PhysRevLett.81.5932, PhysRevLett.96.240501, PhysRevA.79.042340} is decoherence of the qubit caused by the interaction of a photon qubit with the environment, such as in the case of single-mode fiber. Typically, the information is encode into polarization degree of freedom of photon. The interaction of the photons with the birefringent environment of a single-mode fiber makes secure, long-distance communication difficult because the birefringence in fiber randomizes the phase of the photon, leading to a loss of coherence and subsequently loss of information. This effect is called decoherence and it limits the distance over which quantum information can be stored and transmitted and it is one of the obstacle in physical realization of any quantum repeater scheme that uses optical communication of qubit state.

The input polarization state can usually be prepared with high accuracy in a laboratory.  Maintaining that state along the communication channel, however, is much more demanding due to the presence of birefringence. One technique to preserve the coherence of polarization qubit against the detrimental effects of noise encountered in the communication channel is \emph{open-loop} control, where the system is subjected to external, suitably tailored, space-dependent pulses which do not require measurement. This control technique minimizes the undesired interaction of system with the environment if the control pulses are applied faster than correlation length of environment and is called \emph{dynamical decoupling} (DD) \cite{PhysRevA.58.2733,PhysRevLett.82.2417,PhysRevA.70.062310}. The advantage of DD is that it requires modest resources and no additional resource for information encoding, error correction, measurement, or feedback. The physical idea behind this scheme comes from the refocusing technique of nuclear magnetic resonance (NMR), where time dependent, rapid, and strong pulses known as bang-bang controls are applied to the system to suppress decoherence \cite{NMR}.

In the past, techniques based on DD control have been proposed by Wu and Lidar \cite{PhysRevA.70.062310} and others \cite{NJP_2007,PhysRevLett.103.040502,PhysB_2011} for preserving the polarization qubit in an optical fiber, however, no specific scheme has been studied in single-mode fiber that are not polarization preserving. In our previous work, we proposed the Carr-Purcell-Meiboom-Gill (CPMG) DD pulse sequence for preserving polarization qubit in a polarization-maintaining fiber where the birefringence of the fiber was simulated as a Gaussian-distributed, zero-mean
random process \cite{RevSciIns1958}. We numerically showed that effect of decoherence can be minimized with the use of ideal pulses implemented using suitability oriented half-wave plates (HWP) and a fidelity greater than $99\%$ can be achieved \cite{PhysRevA.85.022340}.

Polarization-maintaining fibers are less widely used in telecommunication in comparison to single-mode fiber since they are costly. For this reason, we propose a scheme to preserve the polarization qubit in a single-mode fiber with a more realistic birefringence noise profile, using the Knill dynamical decoupling (KDD) pulse sequence \cite{PhysRevLett.106.240501}. We numerically show that when the self correcting KDD pulse sequence is implemented, using HWPs in a single-mode fiber, then a polarization qubit can be preserved with fidelity greater than $99\%$. In reality, manufacturing a fiber with perfect wave plates is a difficult task. Hence, we also show that in the presence of $0.5\%$ rotation error in a pulse, a fidelity greater than $96\%$ can be achieved. Since any state can be preserved with high fidelity it can be used as delay line or as quantum memory to store photon qubit for short period of time.

We first review, in section \ref{sec:Noise}, the noise in a single-mode fiber and discuss the relevant numerical model. In section \ref{sec:DD}, we discuss DD open-loop control technique and two DD sequences. We finally show in section \ref{sec:Numerical}, through numerical simulation, that with the KDD pulse sequence a polarization qubit can be preserved with $96\%$ fidelity in presence of $0.5\%$ rotation error in HWPs.

\section{Noise in single-mode fibers}
\label{sec:Noise}

Decoherence of the polarization qubit has its origin in optical birefringence. A telecommunication fiber is often called a single-mode fiber, although it supports two orthogonal modes due to its circular symmetry. In a perfect fiber, these two modes have the same phase velocity, but real fibers have some asymmetry due to manufacturing imperfections or stress on the fiber due to deployment in the field. The asymmetry in the fiber breaks the degeneracy of two orthogonal modes, which results in birefringence: a difference in phase velocity of two modes.

Birefringence in an optical fiber can result from both intrinsic and extrinsic perturbations. Intrinsic perturbation can originate during the manufacturing process and is permanent. Form (geometric) birefringence can arise due to non-circular waveguides and stress birefringence is due to forces set up by asymmetry of the core. Extrinsic perturbation can be caused by spooling the fiber or embedding the fiber in the ground, lateral stress, bending, or twisting. These perturbations create linear birefringence and can change as the fiber's external environment changes \cite{BornWolf89,raeySpringer2011}.

For a short section of fiber, birefringence can be considered to be uniform. The difference between the propagation constant of slow and fast modes can be expressed as
\begin{equation}
\Delta\beta =\frac{\omega n_{s}}{c}-\frac{\omega n_{f}}{c}=\frac{\omega \Delta n}{c}
\end{equation}
where $\omega$ is the angular frequency, $c$ is the speed of light, and $\Delta n = n_{s}-n_{f}$ is the differential effective refractive index between slow (s) and fast (f) modes \cite{PMD_Book}.

When a linearly polarized wave (such as $45^{0}$) polarization is launched into the fiber, the asymmetry in the fiber causes a phase shift between the two orthogonal modes. This occurs because these two modes travel with different phase velocities and acquire a relative phase shift that is a function of propagation length in the birefringent fiber. 
As a consequence, the state of polarization (SOP) evolves in a cyclic fashion as the light propagates down the fiber ( i.e., from linear to elliptical to circular and back through elliptical to linear state that is orthogonal to the initial state). The output mode will not be stable as environmental factors such as stress and ambient temperature changes will change the birefringent properties of fiber. This complicates the utilization of single-mode fiber in applications such as quantum cryptography and quantum communication where polarization modes need to be preserved.

\subsection{Noise Model}
\begin{figure}[tb]
\includegraphics[width=8.6cm]{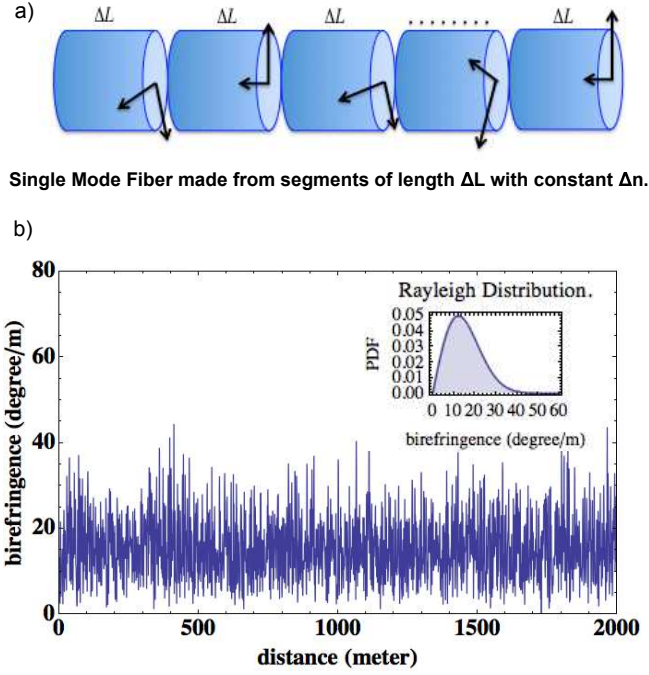}% Here is how to import EPS art
\caption{\label{fig:SM_NOISE} Model of a single-mode optical fiber. a) The fiber is modeled as concatenated segments of fiber of length $\Delta L$ with constant birefringence $\Delta n$. b) Numerically reproduced birefringence in a single-mode fiber, where the scale parameter for Rayleigh distribution is $ \sigma\beta = 12.6 $ degree/m \cite{935820}.}
\end{figure}

Our model of noise closely represents the experimentally measured spatial distribution of birefringence in a single-mode fiber \cite{935820}. It represents the birefringence in different types of fibers: standard step-index and dispersion shifted fibers. In general, axially varying birefringent dephasing in an optical fiber of length $L$ can be represented by a series of concatenated, homogeneous segments of length $\Delta L$ with constant $\Delta n$, as illustrated in Fig. \ref{fig:SM_NOISE}. The birefringence across these segments follows the Rayleigh distribution \cite{Rayleigh}. Here we assume that the fiber only exhibits linear birefrengence. 

The absolute phase difference acquired by the photon after propagating through the $j^{th}$ segment is given by 
\begin{equation}
\delta\phi_{j} =\Delta\beta_{j} \Delta L
\end{equation}
where $\delta\phi$ is the acquired phase after traveling though the $j^{th}$ segment of length $\Delta L$ in the fiber.

These fiber segments together constitute a single phase profile for a particular instance of birefringent noise and corresponding changes in the refractive index $\Delta n$. The output state is obtained by averaging over all phase profiles.

Let us consider the following pure input state:
\begin{equation}
\ket{\psi}= \alpha \ket{H} + \beta \ket{V} .
\end{equation}
For the above input state the density matrix is given by
\begin{equation}
\hat{\rho}_{\mathrm{in}} = \ket{\psi} \bra{\psi} = 
\begin{pmatrix}
\abs{\alpha}^{2} & \alpha \beta^{*} \\
\alpha^{*} \beta & \abs{\beta}^{2}
\end{pmatrix} .
\end{equation}
When the photon travels through the $j^{th}$ segment, the rotation operator $\mathbb{M}_{z}\left(\delta\phi_{j} \right)$ given by
\begin{align}
\mathbb{M}_{z}\left(\delta\phi_{j} \right) & = 
\begin{pmatrix}
\mathrm{e}^{i \frac{\delta\phi_{j} }{2}} & 0 \\
0 & \mathrm{e}^{-\mathrm{i} \frac{\delta\phi_{j} }{2}}
\end{pmatrix} \nonumber \\
&= \cos \left( \frac{\delta\phi_{j}}{2} \right) \mathbb{I} + i \sin \left( \frac{\delta\phi_{j} }{2} \right) \sigma_{z} \nonumber \\
&= \mathrm{e}^{i \frac{\delta\phi_{j} }{2} \sigma_{z}}  \nonumber \\ 
&= \mathbb{R}_{z}\left(\delta\phi_{j} \right)
\end{align}
where $\delta\phi_{j}  = \Delta \beta_{j} \Delta L$ is the phase angle acquired due to propagation through the $j^{th}$ segment of fiber. The rotation operator acts on the photon and rotates the polarization degrees of freedom. The output density matrix is given by
\begin{align}
\hat{\rho}_{j_{\mathrm{out}}} &= \mathbb{M}_{z}\left(\delta\phi_{j} \right) \rho_{in} \mathbb{M}_{z}\left(\delta\phi_{j} \right)^{\dagger} \nonumber \\
&=
\begin{pmatrix}
\abs{\alpha}^{2} & \alpha \beta^{*} \mathrm{e}^{i \delta\phi_{j} }\\
\alpha^{*} \beta \mathrm{e}^{-i \delta\phi_{j} } & \abs{\beta}^{2}
\end{pmatrix} .
\end{align}

After passing through the fiber with $n$ homogeneous concatenated segments the output density matrix is
\begin{align} \label{eq:densityMatrix}
\hat{\rho}_{\mathrm{out}} =
\begin{pmatrix}
\abs{\alpha}^{2} & \alpha \beta^{*} \prod \limits_{j=1}^n \mathrm{e}^{i \delta\phi_{j} }\\
\alpha^{*} \beta \prod \limits_{j=1}^n\mathrm{e}^{-i \delta\phi_{j} } & \abs{\beta}^{2}
\end{pmatrix} .
\end{align}

We model the set of acquired phases $\left\lbrace \delta\phi_{1}, \delta\phi_{2},......,\delta\phi_{n} \right\rbrace$  as random variable  $\hat{\phi}$ with a mean $\left\langle \hat{\phi} \right\rangle = \phi$ and a nonzero variance $\left\langle \Delta\hat{\phi}^2 \right\rangle =  \Delta\phi^2 $.
\begin{align} \label{eq:phase}
\prod \limits_{j=1}^n \mathrm{e}^{\pm i  \delta\phi_{i}} &= \exp \left[ \sum_{j=1}^n \left( \pm i  \delta\phi_{i}\right) \right] \nonumber \\ 
&=\exp \left[ \sum_{j=1}^n \left( \pm i \left\langle \hat{\phi} \right\rangle \pm  i \Delta\hat{\phi} \right) \right]  \nonumber \\
&= \exp \left[ \pm i n \left\langle \hat{\phi} \right\rangle \right] \exp \left[ \pm  i n  \Delta\hat{\phi} \right].
\end{align}

We Taylor expand the factor $\exp \left[ \pm  i n  \Delta\hat{\phi} \right] $  of  Eq. \ref{eq:phase} and take the average to obtain
\begin{align} \label{eq:deco_factor_2}
\left\langle \exp \left[ \pm  i n  \Delta\hat{\phi} \right] \right\rangle &= \left\langle 1 \pm i n  \Delta\hat{\phi} - \frac{1}{2} n^2 \Delta\hat{\phi}^2 + \cdots \right\rangle \nonumber \\
&= 1 \pm i n \left\langle \Delta\hat{\phi} \right\rangle - \frac{1}{2} n^2 \left\langle \Delta\hat{\phi}^2 \right\rangle + \cdots
\end{align}%
Since the mean of variance is zero in  Eq. \ref{eq:deco_factor_2} and average of the variance is $\left\langle \Delta\hat{\phi}^2 \right\rangle =  \Delta\phi^2 $, hence we obtain the expression
\begin{align} \label{eq:decoherence_term}
\left\langle \exp \left[ \pm  i n  \Delta\hat{\phi} \right] \right\rangle &= 1 - \frac{1}{2} n^2  \Delta\phi^2  + \cdots 
& \approx \mathrm{e}^{-\frac{1}{2} n^2 \Delta \phi^2} .
\end{align}%
Using Eq. \ref{eq:decoherence_term}, we can write Eq. \ref{eq:phase} as
\begin{align} \label{approximation}
\left\langle \prod \limits_{j=1}^n \mathrm{e}^{\pm i  \delta\phi_{i}} \right\rangle &= \left\langle \exp \left[ \pm i n \left\langle \hat{\phi} \right\rangle \right]  \right\rangle \left\langle  \exp \left[ \pm  i n  \Delta\hat{\phi} \right] \right\rangle \nonumber \\
&= \mathrm{e}^{ \pm i n \phi } \mathrm{e}^{-\frac{1}{2} n^2 \Delta \phi^2}.
\end{align}%
Using the expression in Eq. \ref{approximation} , the density matrix in Eq. \ref{eq:densityMatrix} can be rewritten as
\begin{align} \label{eq:apprxDensityMatrix}
\hat{\rho}_{\mathrm{out}} =
\begin{pmatrix}
\abs{\alpha}^{2} & \alpha \beta^{*}  \mathrm{e}^{i n \phi } \mathrm{e}^{-\frac{1}{2} n^2 \Delta \phi^2}\\
\alpha^{*} \beta \mathrm{e}^{-i n \phi } \mathrm{e}^{-\frac{1}{2} n^2 \Delta \phi^2} & \abs{\beta}^{2}
\end{pmatrix} .
\end{align}

The state represented by $\hat{\rho}_{\mathrm{out}}$ is no longer pure due to presence of $ \mathrm{e}^{-\frac{1}{2} n^2 \Delta \phi^2} $  in the off-diagonal terms \cite{PhysRevA.88.023857}.

The polarization decoherence discussed above can be inhibited with the use of open-loop quantum control called DD. 

\section{Dynamical Decoupling}
\label{sec:DD}

DD is an open-loop control technique to decouple the system from environmental interaction. One important aspect of this technique is that one can effectively control the dynamical evolution of the system while still eliminating the effects of environment. Even though, environment makes it impossible to achieve an arbitrary unitary evolution of the system, an effective dynamical evolution of the system can still be constructed with this technique. 

This technique is inspired by the nuclear magnetic resonance technique where tailored time dependent perturbations are used to control the system evolution. In DD, we apply sequence of pulses to the system, which is faster than the shortest time scale accessible to the reservoir degree of freedom, such that system-bath coupling is averaged to zero during the evolution. This procedure is conceptually different than NMR, because any decoupling action is only applied to system variable.

To apply the formalism of DD, we consider a system $S$ coupled to a bath $B$, which together form a closed system defined by Hilbert space $\mathcal{\widehat{H}} = \mathcal{\widehat{H}}_{S} \otimes  \mathcal{\widehat{H}}_{B} $, where $\mathcal{\widehat{H}}_{S}$ and $\mathcal{\widehat{H}}_{B}$ denotes the system and bath Hilbert spaces respectively. The overall Hamiltonian is rewritten in an explicit form as %
\begin{equation}
\widehat{H} = \widehat{H}_{S} \otimes\mathbb{I}_{B} + \mathbb{I}_{S} \otimes \widehat{H}_{B} + \widehat{H}_{I}
\end{equation}
where $\mathbb{I}$ is the identity operator and $H_{I}$ is the interaction Hamiltonian which is written as 
\begin{equation}
\widehat{H}_{I} = \mathbb{M}_{z} \otimes \mathbb{B}_{z},
\end{equation}
$\mathbb{M}_{z}$ is the noise operator and $\mathbb{B}_{z}$ is the bath operator that couples the polarization qubit to the noise bath and causes decoherence. To decouple the system from the environment, we introduce time-dependent perturbing Hamiltonian which rotates the qubit around a given axis
on the Bloch sphere \cite{PhysRevLett.82.2417,PhysRevA.85.022340}. The interaction Hamiltonian with added perturbation is given as %
\begin{equation}
\widehat{H}_{I}\left(t \right)= B \left( t \right) \widehat{\sigma}_{z} + f\left( t \right) \widehat{n}\cdot \widehat{\overrightarrow{\sigma}}
\end{equation}
where $B \left( t \right)$ is a scalar function of time and $f\left( t \right)$ is a time-dependent function for the control pulse.

The initial state over $\mathcal{\widehat{H}}$ is represented by the density matrix  $\widehat{\rho}_{\mathrm{tot}}(0) = \widehat{\rho}_{S}(0) \otimes \widehat{\rho}_{B}(0)$, and the state after the coarse-grained dynamics results in open-system evolution defined by  $\widehat{\rho}_{S}(0) \to \widehat{\rho}_{S}(t) = \Tr_{B} \{\widehat{\rho}_{\mathrm{tot}}(t)\}$, where $\Tr_{B}$ denotes the partial trace over $\widehat{H}_{B}$. In the case of the polarization qubit in an optical fiber we assume that the relaxation dynamics for $\widehat{\rho}_{S}(t)$ involves only the decoherence mechanisms represented by coupling operators. The irreversible loss of quantum coherence is caused by presence of $\mathrm{e}^{-\frac{1}{2} n^2 \Delta \phi^2}$ in the off diagonal term of the output density matrix (Eq. \ref{eq:densityMatrix}).

The decoupling field is made to be cyclic and the decoupling operator $\widehat{U}_{1}(t)$ that is periodic over some cycle time $T_{c} > 0$ is given as %
\begin{equation}
\widehat{U}_{1}(t) \equiv  \exp \left\lbrace -i \int_{0}^{t} dt\widehat{H}_{I}(t) \right\rbrace = \widehat{U}_{1}(t+T_{c}) 
\end{equation}

In the interaction picture, the state after the perturbation due to  $\widehat{H}_{I}(t)$ is defined by $\widehat{\rho}_{\mathrm{tot}}(t)=\widehat{U}_{1}(t) \widehat{\tilde{\rho}}_{tot}(t){\widehat{U}}_{1}^\dagger(t)$ and the evolution is defined by a time-varying Hamiltonian, %
\begin{equation}
\widehat{\tilde{H}}(t) = \widehat{U}_{1}^\dagger(t)\widehat{H}_{0}\widehat{U}_{1}(t) .
\end{equation}

The decoupling pulse sequence representing the control field is chosen in such a way that the integral term $\int_{0}^{t} dt\widehat{H}_{I}(t)$ averages to zero and $\widehat{U}_{1}(T_{c}) = \mathbb{I}_{S}$, as a consequence decoherence is minimized.

To preserve the polarization qubit from decoherence, we select the simplest DD pulse sequence which is made up of two $\pi$-pulse separated as shown in Fig. \ref{fig:KDD}. This sequence achieves a fidelity of $99\%$ in preserving the state of the qubit with perfect pulses, but in real world the pulses are never perfect, and  we use a self-correcting pulse sequence known as KDD sequence. This pulse sequence is made up of twenty $\pi$ pulses and it is robust against imperfection.

\subsection{Pulse Imperfection}
The goal of DD is to time reverse the system-bath interaction by applying series of stroboscopic pulses in cycles of period $T_{c}$. If the pulse is instantaneous and perfect, then the dynamics of the system is completely time reversed, and the initial state of the system is recovered. However, if the pulses are finite and have errors, then the system dynamics cannot be completely reversed, and these errors accumulate, and the initial state is lost. In reality, the pulses have finite length and errors and if the number of applied pulses are large and the design of the pulse sequence does not incorporate error compensation, then these errors accumulate and lead to decoherence of the system more severe than due to system environment interaction.

One type of non-ideal pulse has finite length, which imposes limits on the achievable cycle time for the sequence. The constraint on cycle time imposes an upper bound on the maximum achievable DD performance. Another type of non-ideal pulse, which can be more detrimental, is due to rotation error, which is caused by deviation in the control pulse strength. The rotation error can be attributed to rotations about a non ideal axis or deviation in rotation angle.

Several approaches have been proposed \cite{PhysRevLett.95.180501,PhysRevA.75.062310} to make DD insensitive to pulse imperfections and one of the approaches focuses on the design of pulses which are inherently robust or insensitive to imperfect rotation. Rotation about an axis can be achieved either by one single pulse about a given axis or by using a sequence of pulses which are inherently robust against classes of imperfections and generate a rotation close to ideal rotation even in the presence of these errors. In NMR literature, such pulse sequences are called composite pulses \cite{Levitt1996}.

If DD sequence are made up of only $\pi$-rotations, then replacing the $\pi$-rotation with a composite sequence which can achieve a near-ideal rotation in case of imperfection should improve DD performance. KDD sequence uses this design principle and is given as  %
\begin{align}
KDD_{\phi} = & f_{\tau / 2}-(\pi)_{\pi /6 + \phi}-f_{\tau}-(\pi)_{\phi}-f_{\tau}-(\pi)_{\pi /2 + \phi} \nonumber \\
&-f_{\tau}-(\pi)_{\phi}-f_{\tau}-(\pi)_{\pi/6 + \phi}-f_{\tau /2}, 
\end{align}

\begin{figure}[!]
\includegraphics[width=8.6cm]{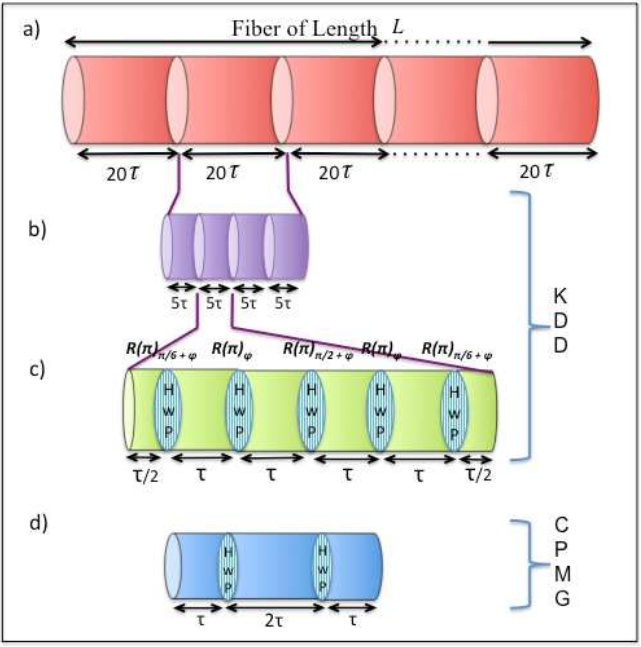}% Here is how to import EPS art
\caption{\label{fig:KDD}A $20$ pulse sequence created by combining 5-pulse block shifted in phase by $\pi/2$. The cyclic repetition of these $20$ pulse sequence is referred to as KDD \cite{PhysRevLett.106.240501}. a) Single-mode fiber of length $L$ divided in to blocks of length $20~\tau$. b) each block is then divided in four parts of length $5~\tau$. c) In each part a KDD block is implemented by introducing $5$ half-wave plate at specified distances. d) CPMG pulse implemented using half-wave plate.}
\end{figure}

where the notation $f_{\tau / 2}$ represents free evolution for a period of $\tau /2$ and $(\pi)_{\pi /6 + \phi}$ is $\pi$ rotation about an axis $\pi /6 + \phi $. It is a 20-pulse, self correcting sequence that is created by combining 5-pulse blocks shifted in phase by $\pi/2$ to form [$KDD_{\phi}-KDD_{\phi+\pi /2}-KDD_{\phi+\pi}-KDD_{\phi+3 \pi /2}$] \cite{PhysRevLett.106.240501}.

We preserve the polarization qubit in a single-mode fiber from birefringent dephasing by implementing the KDD pulse sequence in space by introducing half-wave plate at specified distances in the fiber as shown in Fig. \ref{fig:KDD}.

The impact of pulse imperfections on fidelity in the absence of environment have been analyzed previously and it was found that the KDD sequence is least susceptible to pulse imperfections when compared with other sequences \cite{PhysRevLett.106.240501}.

\section{Numerical Simulation and Results}
\label{sec:Numerical}

\begin{figure*}
\includegraphics{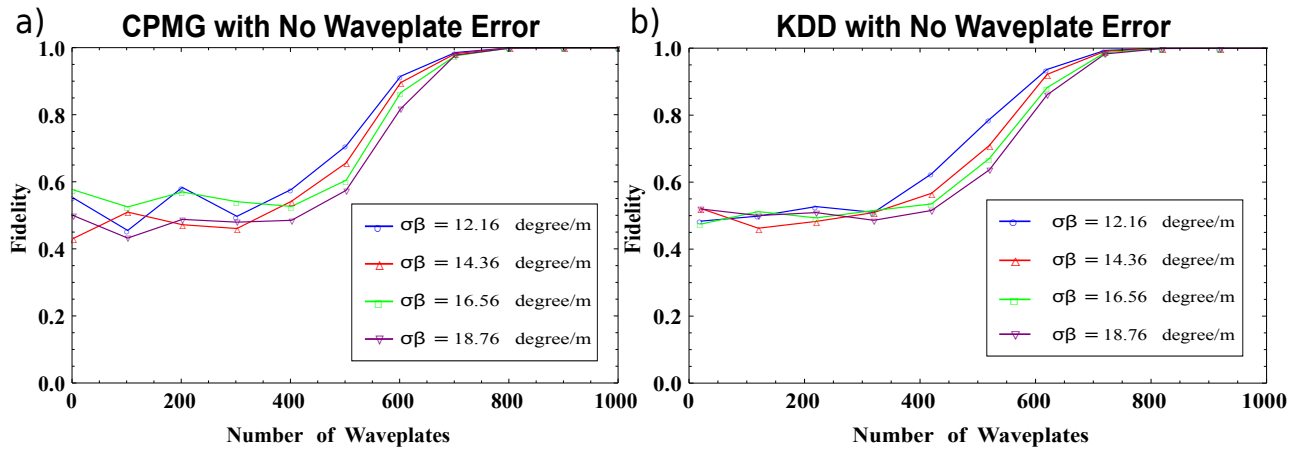}% Here is how to import EPS art
\caption{\label{fig:CPMG_KDD_NoErr} Fidelity of DD sequence in a $500$ m single-mode fiber with perfect pulses. a) CPMG pulse sequence b) KDD pulse sequence.}
\end{figure*}

\begin{figure*}
\includegraphics{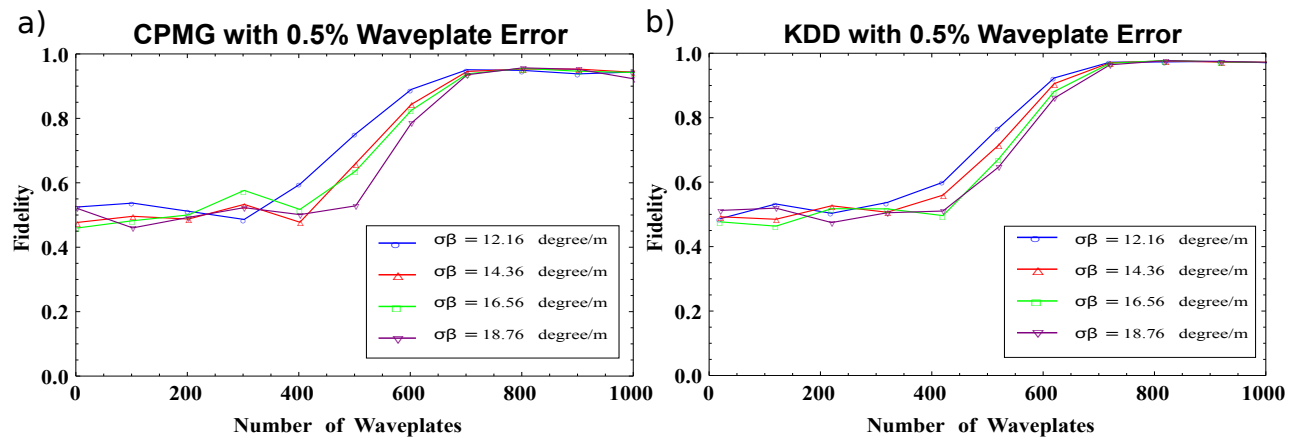}% Here is how to import EPS art
\caption{\label{fig:CPMG_KDD_05p_Err} Fidelity of DD sequence in a $500$ m single-mode fiber with $0.5\%$ pulses error. a) CPMG pulse sequence b) KDD pulse sequence.}
\end{figure*}

We numerically show that state of polarization can be preserved against decoherence caused by the birefringence present in a single-mode fiber by introducing half-wave plates (HWPs) at predetermined positions along the fiber. Since the manufacturing of fibers with waveplates at appropriate positions is limited by manufacturing accuracy, we also show that such intrinsic error can also be minimized when KDD is used. 

To preserve the coherence of the polarization qubit in a single-mode fiber, we use DD pulse sequences implemented with HWPs. We compare the effectiveness of CPMG and KDD pulse sequences in the presence of pulse imperfections to allow for intrinsic errors in the HWP, and we show that as the fiber length increases, the accumulation of errors due to pulse imperfection can be better suppressed with the KDD sequence as compared to the CPMG sequence. 

The birefringence of a single-mode fiber is modeled by generating a set of values according to the Rayleigh distribution, whose probability density function is given as%
\begin{equation}
f(x,\sigma) = \frac{x}{\sigma^2}e^{{-x^2}/{(2 \sigma^2)}},~x \geq 0
\end{equation}
where $\sigma \geq 0$, is the scale parameter of the distribution, and $x$ is the distance along the fiber. Each value is the phase error acquired by the photon as it travels a distance $\Delta L$ along the fiber. A noise profile of the fiber is extrapolated from these phase error values.

We then calculate the fidelity of a specific DD pulse sequence in a fiber of a particular length, number of sections, and initial polarization state. For each section of fiber, the initial state of photon is allowed to freely evolve for a distance according to %
\begin{equation}
\mathbb{M}_{z}\left(\delta\phi_{j}\right)  = 
\begin{pmatrix}
\mathrm{e}^{i \frac{\delta\phi_{j}}{2}} & 0 \\
0 & \mathrm{e}^{-i \frac{\delta\phi_{j}}{2}}
\end{pmatrix},
\end{equation}
where, $\delta\phi_{j}$ includes the phase error from the Rayleigh distribution. The pulse error is then calculated from the normal distribution with zero mean, and the state is rotated according to the particular DD pulse sequence being analyzed. This is repeated for each section in the fiber. We then compare the output state with input state and use fidelity as a measure of effectiveness in preserving the state of photon. Fidelity is defined as%
\begin{equation}
\mathcal{F}= \abs{\bracket{\psi_{i}}{\psi_{f}}}^{2} = \bra{\psi_{i}} \widehat{\rho}_{\mathrm{out}} \ket{\psi_{i}},
\end{equation}
where $\psi_{f}$ and $\psi_{i}$ represent the final and initial state respectively and $\widehat{\rho}_{\mathrm{out}}= \frac{1}{n}\sum{ \ket{\psi_{out}} \bra{\psi_{out}}}$ is average output state over $n$ fiber noise profiles.

To compare the effectiveness of KDD and CPMG pulse sequences in suppressing the decoherence, we first run the simulation for a single-mode fiber of length $500$ meters in absence of pulse error. The result of the simulation is shown in Fig. \ref{fig:CPMG_KDD_NoErr}., where we find that in the absence of pulse errors, the two schemes perform equally well and require 800 wave plates to preserve the polarization qubit in a $500$-meter single-mode fiber.

\begin{figure*}
\includegraphics{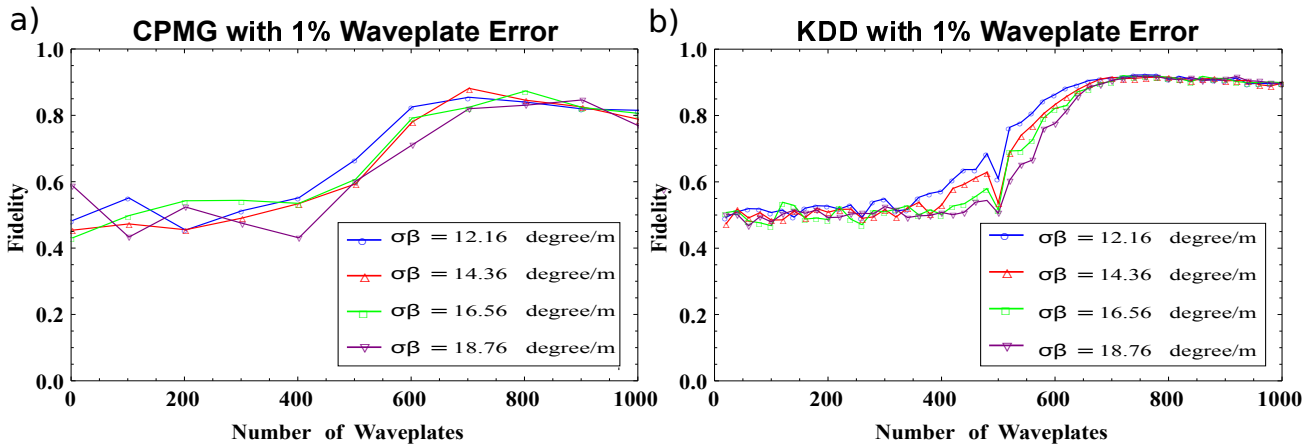}% Here is how to import EPS art
\caption{\label{fig:CPMG_KDD_1p_Err} Fidelity of DD sequence in a $500$ m single-mode fiber with $1\%$ pulses error. a) CPMG pulse sequence b) KDD pulse sequence.}
\end{figure*}

\begin{figure*}
\includegraphics{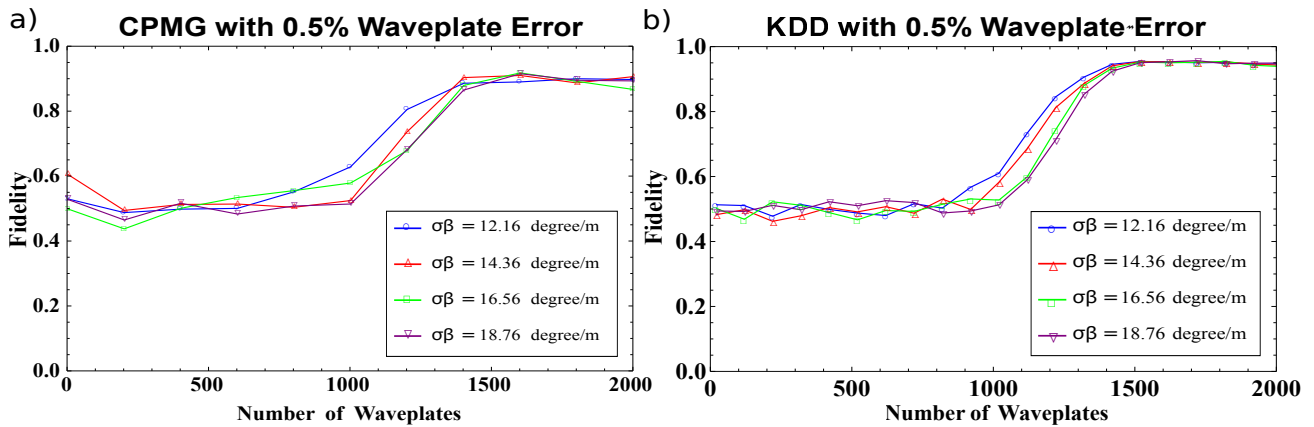}% Here is how to import EPS art
\caption{\label{fig:CPMG_KDD_1km} Fidelity of DD sequence in a $1$ km single-mode fiber with $0.5\%$ pulses error. a) CPMG pulse sequence b) KDD pulse sequence.}
\end{figure*}

We then introduce $0.5\%$ error in the HWPs, distributed as Gaussian with zero mean and run the simulation for a $500$-meter single-mode fiber. The result of the numerical simulation is show in Fig. \ref{fig:CPMG_KDD_05p_Err}. The KDD sequence performs better when compared to CPMG sequence and achieves a fidelity of $96\%$. Furthermore,  when errors in the waveplate is increased from $0.5\%$ to $1\%$, the fidelity of the CPMG sequence drops to $80\%$ due to error accumulation, whereas the fidelity of the KDD sequence remains above $90\%$ due to error cancellation as shown in Fig. \ref{fig:CPMG_KDD_1p_Err} .

 As the fiber length is increased from $500$ meters to one kilometer, the pulse errors accumulate and the fidelity of CPMG falls, whereas the fidelity of the KDD sequence remains the same due to cancellation of errors, as shown in Fig. \ref{fig:CPMG_KDD_1km}.

\section{Conclusion}
We numerically show that horizontal, vertical, and diagonal SOPs can be preserved in a single-mode fiber with the use of the KDD pulse sequence even in the presence of pulse errors, and a fidelity greater than $96\%$ between input and output state can be achieved. This scheme minimizes the dephasing of qubit due to intrinsic and extrinsic perturbation caused during manufacturing and deployment. Wave plates rotate the qubit around the two axes of the Bloch sphere, which averages out the birefringence along the two axes and we expect that polarization mode dispersion will automatically be minimized. Manufacturing and deployment of optical fiber is never error free hence such a scheme is more effective since the pulse sequence is self-correcting. 

Although the results presented here are for diagonal state of polarization but our numerical simulation shows that any qubit state can be preserved with fidelity greater than $99\%$ in absence  of rotation error and with fidelity greater that $96\%$ in presence of rotation error. This scheme has potential applications for physical realization of all optical quantum delay line, quantum memory and quantum repeater schemes.

From an experimental perspective, placement of waveplates at proposed distances can be done during manufacturing process with the help of Bragg transmission grating \cite{doi:10.1117/1.1757455} or controlled twisting of the fiber during manufacturing.

\begin{acknowledgments}
The authors would like to acknowledge the support from the Air Force office of Scientific Research, the Army Research office, and the National Science Foundation.
\end{acknowledgments}

\bibliography{References}

\end{document}